\documentclass[aps,twocolumn,showpacs]{revtex4}
\usepackage{amsmath}
\usepackage{epsfig}

\begin{document}

\title{The explanation of some exotic states in the $cs\bar{c}\bar{s}$ tetraquark system}

\author{Xuejie Liu$^1$}\email[E-mail: ]{1830592517@qq.com}
\author{Hongxia Huang$^2$}
\email[E-mail: ]{hxhuang@njnu.edu.cn (Corresponding author)}
\author{Jialun Ping$^2$}
\email[E-mail: ]{jlping@njnu.edu.cn (Corresponding author)}
\author{Dianyong Chen$^1$}\email[E-mail: ]{chendy@seu.edu.cn}
\author{Xinmei Zhu$^3$}\email[E-mail: ]{zxm_yz@126.com}
\affiliation{$^1$School of Physics, Southeast University, Nanjing 210094, P. R. China}
\affiliation{$^2$Department of Physics, Nanjing Normal University, Nanjing 210023, P.R. China}
\affiliation{$^3$Department of Physics, Yangzhou University, Yangzhou 225009, P. R. China}

\begin{abstract}
Inspired by the recent observation of $\chi_{c0}(3930)$, $X(4685)$ and $X(4630)$ by the LHCb Collaboration and some exotic resonances such as $X(4350)$, $X(4500)$, etc. by several experiment collaborations, the $cs\bar{c}\bar{s}$ tetraquark systems with $IJ^{P}=00^+$, $01^+$ and $02^+$ are systematically investigated in the framework of the quark delocalization color screening model(QDCSM). Two structures, the meson-meson and diquark-antidiquark structures, as well as the channel-coupling of all channels of these two configurations are considered in this work. The numerical results indicate that the molecular bound state $\bar{D}_{s}D_{s}$ with $IJ^{P}=00^+$ can be supposed to explain the $\chi_{c0}(3930)$. Besides, by using the stabilization method, several resonant states are obtained. There are four $IJ^{P}=00^{+}$ states around the resonance mass 4035 MeV, 4385 MeV, 4524 MeV, and 4632 MeV, respectively; one $IJ^{P}=01^{+}$ state around the resonance mass 4327 MeV; and two $IJ^{P}=02^{+}$ states around the resonance mass 4419 MeV and 4526 MeV, respectively. All of them are compact tetraquarks. Among these states, $X(4350)$, $X(4500)$ and $X(4700)$ can be explained as the compact tetraquark state with $IJ^{P}=00^{+}$, and the $X(4274)$ is possible to be a candidate of the compact tetraquark state with $IJ^{P}=01^{+}$. More experimental tests are expected to check the existence of all these possible resonance states.
\end{abstract}

\maketitle

\setcounter{totalnumber}{5}

\section{\label{sec:introduction}Introduction}

The underlying theory of strong interaction is quantum chromodynamics (QCD), which is nonperturbative at low energy. Therefore, it is very difficult to be solved by the hadron spectrum model independently. Although the traditional quark model can well describe the hadron spectrum, which is classified as mesons, composed of $q\bar{q}$ and baryons, composed of $qqq$~\cite{tr_model1,tr_model2}, the emergence of plenty of states or resonant structures in experiments can not fit the hadron spectrum predicted by the naive quark model in last two decades. Those states are called exotic states. A lot of effort has been made to understand the nature of these states, but there is still controversy about their nature.

Actually, most of these states are in the proximity of di-hadron thresholds. To name a few famous examples,
the observed $X(3872)$~\cite{Bell-X(3872)} and $Z_{c}(3900)$~\cite{Zc(3900)1,Zc(3900)2,Zc(3900)3} are around the $D\bar{D}^{*}$ threshold and the $Z_{c}(4020)$~\cite{Zc(4020)1,Zc(4020)2} are near the $D^{*}\bar{D}^{*}$ threshold, etc. Recently, more representative states were searched for experimentally, such as $Z_{c}(3985)$~\cite{Zc(3985)}, which are near the $\bar{D}_{s}D^{*}$ and $\bar{D}^{*}_{s}D$ thresholds.

Inspired by the results of recent LHCb experiment about the $X_{c0}(3930)$ which is just below the $D_{s}\bar{D}_{s}$ threshold~\cite{LHCb_X3930.1,LHCb_X3930.2}, we have a great interest in studying the tetraquark system composed of $cs\bar{c}\bar{s}$. Besides, several experiment collaborations have discovered some relevant resonance states. For instance, in 2009, the CDF Collaboration reported a new state $X(4140)$ in the $J/\psi\phi$ invariant mass distribution~\cite{CDF-CCSS} and this structure was then observed by other collaborations, such as LHCb, CMS, D0, and BABAR in next few years~\cite{LHCb-CCSS,CMS-CCSS,D0-CCSS,BABAR-CCSS}; in 2010, a narrow resonance $X(4350)$ was found in the $\gamma\gamma\longrightarrow J/\psi\phi$ process by the Belle Collaboration~\cite{Bell-CCSS}, which favored the spin parity $J^{PC}=0^{++}$ or $2^{++}$; in 2017, the CDF Collaboration observed the resonance state $X(4274)$ in the $B^+\longrightarrow J/\psi \phi K^+$ decay with $3.1\sigma$ significance~\cite{CDF-CCSS-4274}; in 2016, in the $B^+\longrightarrow J/\psi \phi K^+$ decay, the LHCb Collaboration confirmed the existence of the $X(4140)$ and $X(4274)$~\cite{LHCb-CCSS-2017.1}. Their quantum numbers are measured to be $J^{PC}=1^{++}$; in the same process, the Collaboration observed two higher resonances, $X(4500)$ and $X(4700)$ with $J^{PC}=0^{++}$~\cite{LHCb-CCSS-2017.2}; in 2021, an improved full amplitude analysis of the $B^+\longrightarrow J/\psi \phi K^+$ decay is performed by using 6 times larger signal yield than the previous analysis, the LHCb Collaboration discovered two new hadron states, which are $X(4685)$ and $X(4630)$~\cite{LHCb-CCSS-2021}.

To identify the internal structure of these resonance states, there are a large number of theoretical studies about the tetraquark state $cs\bar{c}\bar{s}$.  In Ref~\cite{Canonical-CCSS}, the $X(4140)$ resonance appeared as a cusp in the $J/\psi\phi$ channel and the $X(4274)$, $X(4500)$, $X(4700)$ were all defined as conventional charmonium states by using a nonrelativistic constituent quark model, respectively. In the relativized quark model, the resonance of $X(4140)$ was regarded as the $cs\bar{c}\bar{s}$ tetraquark ground state, the $X(4700)$ was assigned as the 2S excited tetraquark state, $X(4500)$ was explained as the tetraquark composed of one 2S scalar diquark and one scalar antidiquark, and $X(4274)$ was a good candidate of the conventional $\chi_{c1}$ state~\cite{relative-CCSS}. It was argued that the cusp effects might explain the structure of the $X(4140)$, but failed to account for the $X(4274)$~\cite{effect-cups-CCSS}. In Ref~\cite{di-antiquark-CCSS}, the $X(4140)$ and $X(4270)$ with the tetraquark interpretation was consistent with $X(4350)$ while the interpretation of the $X(4500)$ and $X(4700)$ needed orbital or radial excitation in the simple color-magnetic interaction model. Maiani et al. proposed that the $X(4140)$ and the $X(4274)$ could be explained as the ground state 1S-multiplet of diquark-antiquark tetraquarks while the $X(4500)$ and $X(4700)$ were radially excited 2S states~\cite{di-antiquark-CCSS1}. Stancu argued that $X(4140)$ could be the strange partner of X(3872) in a tetraquark interpretation within a simple quark model with the chromomagnetic interaction~\cite{Stancu-CCSS}. In Ref~\cite{QCD-sum-rule-CCSS}, based on the diquark-antidiquark configuration within the framework of QCD sum rules, the $X(4500)$ and the $X(4700)$ were interpreted as the D-wave $cs\bar{c}\bar{s}$ tetraquark states of $J^{P}=0^+$. According to the calculation with multiquark color flux-tube model, Deng et al. pointed out that the $X(4500)$ and the $X(4700)$ were S-wave radial excited states $[cs][\bar{c}\bar{s}]$~\cite{color-flux-tube-CCSS}. Moreover, Yang explained the $X(4274)$ as the $cs\bar{c}\bar{s}$ tetraquark states with $J^{PC}=1^{++}$, $X(4350)$ as a good candidate of the compact tetraquark state with $J^{PC}=0^{++}$, and the $X(4700)$ as the 2S radial excited tetraquark state with $J^{PC}=0^{++}$~\cite{CHQM-CCSS}. In Refs.~\cite{wang-diquark-CCSS1,wang-diquark-CCSS2}, the $X(4500)$ was observed as the first radial excited state of the axial-vector-diquark-axial-vector-antidiquark type scalar $cs\bar{c}\bar{s}$ tetraquark state and the $X(4700)$ was assigned as the ground state vector-diquark-vector-antidiquark type scalar $cs\bar{c}\bar{s}$ tetraquark state, but the results disfavored assigning the $X(4140)$ to the $J^{PC}=1^{++}$ diquark-antiquark type $cs\bar{c}\bar{s}$ tetraquark state. A rescattering mechanism was used in Ref~\cite{rescattering-CCSS} to understand the nature of $X(4140)$, $X(4350)$, $X(4500)$ and $X(4700)$, among which the $X(4140)$ and the $X(4700)$ could be simulated due to the $D^{*}_{s}D_{s}$ rescattering and the $\psi^{'}\phi$ rescattering. However, this mechanism failed to generate the $X(4274)$ and $X(4500)$, which lead to the proposal that they might be the genuine resonances. In Ref~\cite{Ebert}, the masses of the excited heavy tetraquarks with hidden charm were calculated
within the relativistic diquark-antidiquark picture, and the results showed that $X(3872)$, $Y(4260)$, $Y(4360)$, $Z(4248)$, $Z(4433)$ and $Y(4660)$ could be tetraquark states with hidden charm.

In this work, to see whether these exotic resonances can be described by $cs\bar{c}\bar{s}$ tetraquark systems with $J^{P}=0^{+}, 1^{+}$ and $2^{+}$, we systematically study the properties of these exotic resonances by using the quark delocalization color screening model (QDCSM)~\cite{QDCSM_explain1}, which was proposed particularly to study the similarities between nuclear and molecular forces. According to the characteristics of QDCSM, it can give a good description of the properties of the deuteron, nucleon-nucleon, and hyperon-nucleon interactions~\cite{QDCSM_explain2, QDCSM_explain3}. In the present calculation, two configurations, the meson-meson ($q\bar{q}-q\bar{q}$) and the diquark-antidiquark ($qq-\bar{q}\bar{q}$), are taken into account. Besides, to be more convincing, the channel coupling effect of $cs\bar{c}\bar{s}$ tetraquark systems is also included.

This work is organized as follows. In section~\ref{model}, we present a review of the quark delocalization color screening model and the wave functions of the total system in the present work. The numerical results and a discussion for the tetraquarks are given in Section~\ref{results}. Finally, the last section is devoted to a brief summary.

\section{THE QUARK DELOCALIZATION COLOR SCREENING MODEL (QDCSM) AND WAVE FUNCTIONS }{\label{model}}
\subsection{The quark delocalization color screening model (QDCSM)}
The quark delocalization color screening model (QDCSM) is an extension of the native quark cluster model~\cite{native} and was developed with aim of
addressing multiquark systems. The detail of QDCSM can be found in the Refs.~\cite{QDCSM_explain1, QDCSM1, QDCSM2}.
Here, the general form of the four body complex Hamiltonian is given by
\begin{equation}
H = \sum_{i=1}^{4} \left(m_i+\frac{\boldsymbol{p}_i^2}{2m_i}\right)-T_{CM}+\sum_{j>i=1}^4V(r_{ij}),\\
\end{equation}
where the center-of-mass kinetic energy, $T_{CM}$ is subtracted without losing generality since we mainly focus on the internal relative motions of the multiquark system. The interplay is of two body potential which includes color-confining, $V_{CON}$, one-gluon exchange, $V_{OGE}$, and Goldstone-boson exchange, $V_{\chi}$, respectively,
\begin{equation}
V(r_{ij}) = V_{CON}(r_{ij})+V_{OGE}(r_{ij})+V_{\chi}(r_{ij})
\end{equation}
In this work, we focus on the low-lying positive parity $cs \bar{c} \bar{s}$ tetraquark states of $s-$wave, and the spin-orbit and tensor interactions are not included. The potential $V_{OGE}(r_{ij})$ can be written as
\begin{equation}
V_{OGE}(r_{ij}) = \frac{1}{4}\alpha_s \boldsymbol{\lambda}^{c}_i \cdot
\boldsymbol{\lambda}^{c}_j
\left[\frac{1}{r_{ij}}-\frac{\pi}{2}\delta(\boldsymbol{r}_{ij})(\frac{1}{m^2_i}+\frac{1}{m^2_j}
+\frac{4\boldsymbol{\sigma}_i\cdot\boldsymbol{\sigma}_j}{3m_im_j})\right]
\end{equation}
where $m_{i}$ and $\boldsymbol{\sigma}$ are the quark mass and the Pauli matrices, respectively. The $\boldsymbol{\lambda^{c}}$ is SU(3) color matrix. The QCD-inspired effective scale-dependent strong coupling constant, $\alpha_s^{ij}$, offers a consistent description of mesons from light to heavy quark sector.

Similary, the confining interaction $V_{CON}(r_{ij})$ can be expressed as
\begin{equation}
 V_{CON}(r_{ij}) =  -a_{c}\boldsymbol{\lambda^{c}_{i}\cdot\lambda^{c}_{j}}[f(r_{ij})+V_{0_{ij}}],
\end{equation}
and the $f(r_{ij})$ can be written as
\begin{equation}
 f(r_{ij}) =  \left\{ \begin{array}{ll}r_{ij}^2 &\qquad \mbox{if }i,j\mbox{ occur in the same cluster} \\
\frac{1 - e^{-\mu_{ij} r_{ij}^2} }{\mu_{ij}} & \qquad \mbox{if }i,j\mbox{ occur in different cluster} \\
\end{array} \right.
\end{equation}
where the color screening parameter $\mu_{ij}$ is determined by fitting the deuteron properties, $NN$ and $NY$ scattering phase shifts, with $\mu_{qq}= 0.45$, $\mu_{qs}= 0.19$
and $\mu_{ss}= 0.08$, satisfying the relation $\mu_{qs}^{2}=\mu_{qq}\mu_{ss}$, where $q$ represents $u$ or $d$ quark. When extending to the heavy-quark case, we found that the dependence of the parameter $\mu_{cc}$ is not very significant in the calculation of the $P_{c}$ states~\cite{Pc_huang1} by taking it from $0.0001$ to $0.01$.
So here we take $\mu_{cc}=0.01$. Then $\mu_{sc}$ and $\mu_{uc}$ are obtained by the relation $\mu^{2}=\mu_{ss}\mu_{cc} $ and $\mu^{2}=\mu_{uu}\mu_{cc}$, respectively.

The Goldstone-boson exchange interactions between light quarks appear because the dynamical breaking of chiral symmetry. For the $cs\bar{c}\bar{s}$ system, the $\pi$ and $K$ exchange interactions do not appear because there is no up or down quarks herein. Only the following $\eta$ exchange term works between the $s\bar{s}$ pair.
\begin{equation}
V_{\chi}(r_{ij})  =  v^{\eta}_{ij}\left[\left(\lambda _{i}^{8}\cdot
\lambda _{j}^{8}\right)\cos\theta_P-(\lambda _{i}^{0}\cdot
\lambda_{j}^{0}) \sin\theta_P\right] \label{sala-Vchi1}
\end{equation}
with
\begin{eqnarray}
\nonumber
v^{\eta}_{ij} &=&  {\frac{g_{ch}^{2}}{{4\pi}}}{\frac{m_{\chi}^{2}}{{\
12m_{i}m_{j}}}}{\frac{\Lambda _{\chi}^{2}}{{\Lambda _{\chi}^{2}-m_{\chi}^{2}}}}
m_{\chi}                                 \\
&&\left\{(\boldsymbol{\sigma}_{i}\cdot\boldsymbol{\sigma}_{j})
\left[ Y(m_{\chi}\,r_{ij})-{\frac{\Lambda_{\chi}^{3}}{m_{\chi}^{3}}}
Y(\Lambda _{\chi}\,r_{ij})\right] \right\}
\end{eqnarray}
where $Y(x)=e^{-x}/x$ is the standard Yukawa function. The $\boldsymbol{\lambda^{a}}$ is the SU(3) flavor Gell-Mann matrix. The mass of the $\eta$ meson is taken from the experimental value~\cite{PDG}. Finally, the chair coupling constant, $g_{ch}$, is determined from the $\pi NN$ coupling constant through
\begin{equation}
\frac{g_{ch}^{2}}{4\pi}=\left(\frac{3}{5}\right)^{2} \frac{g_{\pi NN}^{2}}{4\pi} {\frac{m_{u,d}^{2}}{m_{N}^{2}}}
\end{equation}
which assumes that flavor SU(3) is an exact symmetry, only broken by the different mass of the strange quark. The model parameters and the masses of the ground mesons are listed in Tables~\ref{parameters} and \ref{mass}, respectively.

\begin{table}[ht]
\caption{\label{biaoge}Model parameters. The masses of mesons take their experimental values.
$m_{\eta}=2.77$ fm$^{-1}$.}
\begin{tabular}{cccccccc}
 \hline \hline
Quark masses  &$m_u$(MeV)                       & 313 \\
              &$m_s$(MeV)                       & 536 \\
              &$m_c$(MeV)                       & 1728 \\
confinement   &$b(fm)$                          &0.3\\
              &$a_{c}$(MeV $fm^{-2}$)           &101 \\
              &$V_{0_{uu}}$(MeV)                &-2.2543\\
              &$V_{0_{us}}$(MeV)                &-1.7984\\
              &$V_{0_{uc}}$(MeV)                &-1.3231\\
              &$V_{0_{ss}}$(MeV)                &-1.3649\\
              &$V_{0_{sc}}$(MeV)                &-0.6739\\
              &$V_{0_{cc}}$(MeV)                &0.7555\\
OGE           &$\alpha_{s}^{uu}$                &0.2567\\
              &$\alpha_{s}^{us}$                &0.2970\\
              &$\alpha_{s}^{uc}$                &0.3805\\
              &$\alpha_{s}^{ss}$                &0.1905\\
              &$\alpha_{s}^{sc}$                &0.6608\\
              &$\alpha_{s}^{cc}$                &1.6717\\
\hline\hline
\end{tabular}
\label{parameters}
\end{table}

\begin{table}[ht]
\caption{The Masses (in MeV) of the ground mesons. Experimental values are taken
from the Particle Data Group (PDG)~\cite{PDG}.}
\begin{tabular}{lcccccccc}
\hline \hline
~~~~~~&~~$K$~~  &~~$K^{*}$~~  &~~$\pi$~~ &~~$\rho$~~ &~~$\eta_{s\bar s}$~~ &~~$\phi$~~~~~~ \\ \hline
Expt  &495      & 892         &139       &770        &958                  &1020     \\
Model &495      & 892         &139       &770        &958                  &1020     \\
~~~~~~&~~$D_{s}$~~  &~~$D_{s}^{*}$~~  &~~$\eta_{c\bar c}$~~ &~~$J/\psi$~~ &~~$D$~~ &~~$D^{*}$~~~~~~ \\
Expt  &1968         &2112             &2983                 &3096         &1865    &2007         \\
Model &1968         &2112             &2983                 &3096         &1865    &2007       \\
\hline\hline
\end{tabular}
\label{mass}
\end{table}

In QDCSM, the quark delocalization is realized by specifying the single particle orbital
wave function of QDCSM as a linear combination of left and right Gaussian, the single
particle orbital wave functions used in the ordinary quark cluster model,
\begin{eqnarray}
\psi_{r}(\boldsymbol{r},\boldsymbol{s}_{i},\epsilon)&=&(\phi_{R}(\boldsymbol{r},\boldsymbol{s}_{i})
  +\epsilon\phi_{L}(\boldsymbol{r},\boldsymbol{s}_{i}))/N(\epsilon), \label{rl1} \\
\psi_{l}(\boldsymbol{r},\boldsymbol{s}_{i},\epsilon)&=&(\phi_{L}(\boldsymbol{r},\boldsymbol{s}_{i})
  +\epsilon\phi_{R}(\boldsymbol{r},\boldsymbol{s}_{i}))/N(\epsilon), \label{rl2} \\
N(\epsilon)&=& \sqrt{1+\epsilon^2+2\epsilon e^{-s^2_{i}/{4b^2}}},\\
\phi_{R}(\boldsymbol{r},\boldsymbol{s}_{i})&=&(\frac{1}{\pi b^2})^{\frac{3}{4}}
e^{-\frac{1}{2b^2}(\boldsymbol{r}-\frac{2}{5}s_{i})^2},\\
\phi_{L}(\boldsymbol{r},\boldsymbol{s}_{i})&=&(\frac{1}{\pi b^2})^{\frac{3}{4}}
e^{-\frac{1}{2b^2}(\boldsymbol{r}+\frac{3}{5}s_{i})^2},
\end{eqnarray}
The $\boldsymbol{s}_{i}$, $i=1,2,..., n$, are the generating coordinates, which are introduced to
expand the relative motion wave function~\cite{si_1,si_2,si_3}. The mixing parameter
$\epsilon(s_{i})$ is not an adjusted one but determined variationally by the dynamics of the
multi-quark system itself. This assumption allows the multi-quark system to choose its
favorable configuration in the interacting process. It has been used to explain the cross-over
transition between the hadron phase and the quark-gluon plasma phase~\cite{si_4}.

\subsection{The wave function}
\begin{figure*}[!htb]
\includegraphics[scale=0.25]{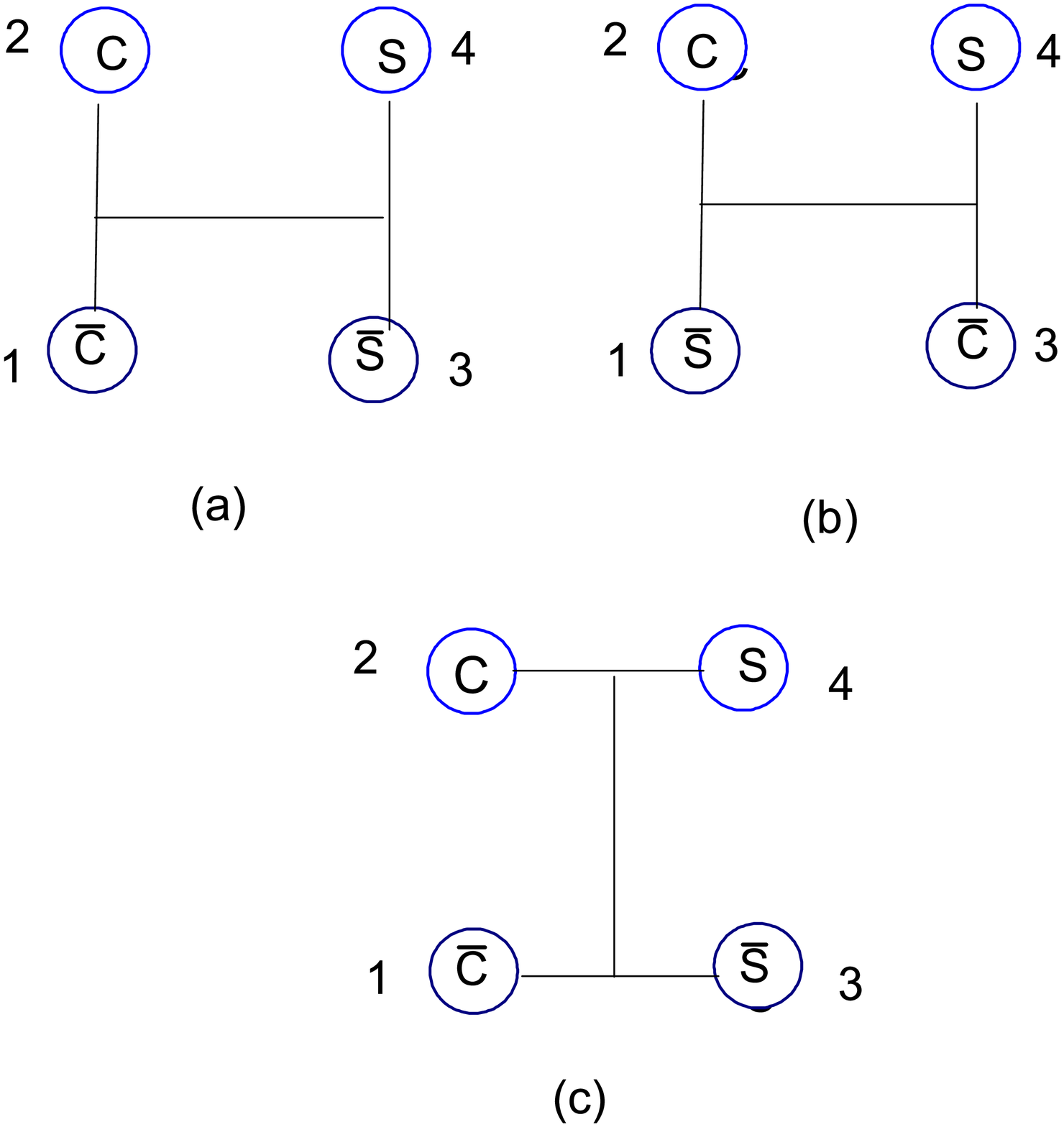}
\vspace{-2.5cm} \caption{Two types of configurations in $c\bar c s\bar s$ tetraquarks. Panel(a) and panel (b) is the meson-meson configuration, panel(c) is diquark-antidiquark. }
\label{fig1}
\end{figure*}

In this work, we focus on the double heavy $c\bar c s\bar s$ system by using the resonance group method~\cite{RGM}.
Figure~\ref{fig1} shows two kinds of configurations for this system, which are the meson-meson structures shown in Fig.~\ref{fig1} (a) and (b), and the diquark-antidiquark structure shown in Fig.~\ref{fig1}(c). For the purpose of solving a manageable 4-body problem, currently, the system calculation considers only these two structures. But an economic way is used to combine these two configurations to see the effect of the multi-channel coupling.
Four fundamental degrees of freedom, which are color, spin, flavor, and orbit are generally accepted by the QCD theory at the quark level. The multiquark system's wave function is an internal product of the color, spin, flavor, and orbit terms.

\subsubsection{The color wave function}
Plenty of color structures in multiquark systems will be available with respect those of conventional hadrons such as $q\bar{q}$ mesons and $qqq$ baryons. In this section, the goal is to construct the colorless wave function of a 4-quark system. For the meson-meson configurations, the color wave functions of a $q\bar{q}$ cluster are listed.
\begin{eqnarray}
\nonumber
C^{1}_{[111]} &=& \sqrt{\frac{1}{3}}(r\bar{r}+g\bar{g}+b\bar{b}) \\ \nonumber
C^{2}_{[21]} &=& r\bar{b}, C^{3}_{[21]} =  -r\bar{g}                    \\ \nonumber
C^{4}_{[21]} &=& g\bar{b}, C^{5}_{[21]} =  -b\bar{g}              \\ \nonumber
C^{6}_{[21]} &=& g\bar{r}, C^{7}_{[21]} =   b\bar{r}         \\ \nonumber
C^{8}_{[21]} &=&  \sqrt{\frac{1}{2}}(r\bar{r}-g\bar{g})       \\ \nonumber
C^{9}_{[21]} &=&  \sqrt{\frac{1}{6}}(-r\bar{r}-g\bar{g}+2b\bar{b}) \\
\end{eqnarray}

where the subscript [111] and [21] stand for color-singlet($\textbf{1}_{c}$) and color-octet($\textbf{8}_{c}$), respectively. So, the $SU(3)_{color}$ wave functions of color-singlet (two color-singlet cluters, $\textbf{1}_{c}\otimes\textbf{1}_{c}$) and hidden-color (two color-octet clusters, $\textbf{8}_{c}\otimes\textbf{8}_{c}$) channels are given respectively.
\begin{equation}
\chi^{c}_{1} = C^{1}_{[111]}C^{1}_{[111]}
\end{equation}
\begin{equation}
\begin{split}
  \chi^{c}_{2} =&\sqrt{\frac{1}{8}}(C^{2}_{[21]}C^{7}_{[21]}-C^{4}_{[21]}C^{5}_{[21]}-C^{3}_{[21]}C^{6}_{[21]}\\
                &+C^{8}_{[21]}C^{8}_{[21]}-C^{6}_{[21]}C^{3}_{[21]}+C^{9}_{[21]}C^{9}_{[21]}\\
                &-C^{5}_{[21]}C^{4}_{[21]}+C^{7}_{[21]}C^{2}_{[21]})
\end{split}
\end{equation}

For the diquark-antidiquark structure, the color wave functions of the diquark clusters is given,
\begin{eqnarray}
\nonumber
  C^{1}_{[2]} &=& rr,  C^{2}_{[2]} = \sqrt{\frac{1}{2}}(rg+gr) \\ \nonumber
  C^{3}_{[2]} &=& gg,  C^{4}_{[2]} = \sqrt{\frac{1}{2}}(rb+br)\\ \nonumber
  C^{5}_{[2]} &=& \sqrt{\frac{1}{2}}(gb+bg),  C^{6}_{[2]} = bb \\ \nonumber
  C^{7}_{[11]} &=& \sqrt{\frac{1}{2}}(rg-gr),  C^{8}_{[11]} = \sqrt{\frac{1}{2}}(rb-br)\\ \nonumber
  C^{9}_{[11]} &=&\sqrt{\frac{1}{2}}(gb-bg) \\
\end{eqnarray}
And the color wave functions of the antidiquark clusters can be writen as:
\begin{eqnarray}
\nonumber
  C^{1}_{[22]} &=& \bar{r}\bar{r},  C^{2}_{[22]} = -\sqrt{\frac{1}{2}}(\bar{r}\bar{g}+\bar{g}\bar{r})\\  \nonumber
  C^{3}_{[22]} &=& \bar{g}\bar{g},  C^{4}_{[22]} = \sqrt{\frac{1}{2}}(\bar{r}\bar{b}+\bar{b}\bar{r}) \\  \nonumber
  C^{5}_{[22]} &=& -\sqrt{\frac{1}{2}}(\bar{g}\bar{b}+\bar{b}\bar{g}), C^{6}_{[22]} = \bar{b}\bar{b} \\ \nonumber
  C^{7}_{[211]}&=& \sqrt{\frac{1}{2}}(\bar{r}\bar{g}-\bar{g}\bar{r}), C^{8}_{[211]} = -\sqrt{\frac{1}{2}}(\bar{r}\bar{b}-\bar{b}\bar{r}) \\ \nonumber
  C^{9}_{[211]} &=& \sqrt{\frac{1}{2}}(\bar{g}\bar{b}-\bar{b}\bar{g}) \\
\end{eqnarray}
The color wave functions of the diquark-antidiquark structure shown in Fig.~\ref{fig1}(c) are $\chi^{c}_{3}$ (color sextet-antisextet clusters, $\textbf{6}_{c}\otimes\bar{\textbf{6}}_{c}$) and $\chi^{c}_{4}$ (color-triplet-antitriplet cluster, $\textbf{3}_{c}\otimes\bar{\textbf{3}}_{c}$).
\begin{equation}
\begin{split}
\chi^{c}_{3} = &\sqrt{\frac{1}{6}}(C^{1}_{[2]}C^{1}_{[22]}-C^{2}_{[2]}C^{[2]}_{[22]}+C^{3}_{[2]}C^{3}_{[22]} \\
               &+C^{4}_{[2]}C^{4}_{[22]}-C^{5}_{[2]}C^{5}_{[22]}+C^{6}_{2}C^{6}_{22})
\end{split}
\end{equation}

\begin{equation}
\begin{split}
  \chi^{c}_{4} =&\sqrt{\frac{1}{3}}(C^{7}_{[11]}C^{7}_{[211]}-C^{8}_{[11]}C^{8}_{[211]}+C^{9}_{[11]}C^{9}_{[211]})
\end{split}
\end{equation}

\subsubsection{The flavor wave function}
For the flavor degree of freedom, since the quark content of the tetraquark systems are two heavy quarks and two strange quarks, the isoscalar sector is $I=0$. The flavor wave functions denoted as $F^{i}_{I}$, with the superscript $I$ referring to isoscalar, can be written as
\begin{eqnarray}
  \nonumber  F^{1}_{0}&=& c\bar{c}s\bar{s}  \\
  \nonumber  F^{2}_{0}&=& c\bar{s}s\bar{c}  \\
             F^{3}_{0}&=& cs\bar{c}\bar{s}
\end{eqnarray}

\subsubsection{The spin wave function}
For the spin, the total spin $S$ of tetraquark states ranges from 0 to 2. All of them are considered.
The wave functions of two body clusters are
\begin{eqnarray}
\nonumber \chi_{11}&=& \alpha\alpha,\\
\nonumber \chi_{10} &=& \sqrt{\frac{1}{2}}(\alpha\beta+\beta\alpha)\\
\nonumber \chi_{1-1} &=& \beta\beta \\
            \chi_{00} &=& \sqrt{\frac{1}{2}}(\alpha\beta-\beta\alpha)
\end{eqnarray}

Then, the total spin wave functions $S^{i}_{s}$ are obtained by considering the coupling of two subcluster spin wave functions with SU(2) algebra, and the total spin wave functions of four-quark states can be read as
\begin{eqnarray}
\nonumber S^{1}_{0}&=&\chi_{00}\chi_{00}\\
\nonumber S^{2}_{0}&=&\sqrt{\frac{1}{3}}(\chi_{11}\chi_{1-1}-\chi_{10}\chi_{10}+\chi_{1-1}\chi_{11})\\
\nonumber S^{3}_{1}&=&\chi_{00}\chi_{11}\\
\nonumber S^{4}_{1}&=&\chi_{11}\chi_{00}\\
\nonumber S^{5}_{1}&=&\sqrt{\frac{1}{2}}(\chi_{11}\chi_{10}-\chi_{10}\chi_{11}) \\
S^{6}_{2}&=&\chi_{11}\chi_{11}
\end{eqnarray}

\subsubsection{The orbital wave function}
Among the different methods to solve the Schr\"{o}dinger-like 4-body bound state equation, we use the resonating group method (RGM)~\cite{RGM}, which is one of the most extend tools to solve eigenvalue problems and scattering problems. The total orbital wave functions can be constructed by coupling the orbital wave function of two internal cluster and the relative motion wave function between two clusters.
\begin{equation}
\psi^{L}=\psi_{1}(R_{1})\psi_{2}(R_{2})\chi_{L}(R)
\end{equation}
where $R_{1}$ and $R_{2}$ are the internal coordinates for the cluster 1 and cluster 2. $R=R_{1}-R_{2}$ is the relative coordinate between the two clusters 1 and 2. The $\psi_{1}$ and $\psi_{2}$ are the internal cluster orbital functions of the clusters 1 and clusters 2, and $\chi_{L}(R)$ is the relative motion wave function between two clusters, which is expanded by the gaussian bases
\begin{eqnarray}
\begin{split}
\chi_{L}(R)=&\sqrt{\frac{1}{4\pi}}(\frac{3}{2\pi b^2})\sum^{n}_{i=1}C_{i} \\
   &\times \int{exp}[-\frac{3}{4b^2}(R-s_{i})^2]Y_{LM}(\hat{s_{i}})d\hat{s_{i}}
\end{split}
\end{eqnarray}
where n is the number of gaussian bases, which is determined by the stability of the results.

Finally, to fulfill the Pauli principle, the complete wave function is written as
 \begin{equation}
   \psi=A[[\psi^{L}S^{j}_{s}]_{JM_{J}}F^{i}_{I}\chi^{c}_{k}]
 \end{equation}
where A is the antisymmetry operator of double-heavy tetraquarks.
In this work, the operator A is defined as $A=1$ due to the absence of any homogeneous quarks in the $c\bar{c}s\bar{c}$ system.

\section{RESULTS AND DISCUSSIONS}{\label{results}}
The low-lying $S-$wave states of $c\bar{c}s\bar{s}$ tetraquark are systematically investigated herein. The parity for $c\bar{c}s\bar{s}$ tetraquark is positive under our assumption that the total orbital angular momenta $L$ is 0. Accordingly, the total angular momenta, $J$, can take values of 0, 1, 2.
The value of isospin can only be 0 for the $c\bar{c}s\bar{s}$ tetraquark system. Two structures of $c\bar{c}s\bar{s}$ tetraquark, meson-meson and diquark-antidiquark structures, are investigated. In each structure, all possible states are considered, which are listed in Table~\ref{channels}. The $F^{i}_{I}; S^{j}_{s}; \chi^{c}_{k}$ shows the necessary basis combination in flavor $(F^{i}_{I})$, spin $(S^{j}_{s})$
and color $(\chi^{c}_{k})$ degrees of freedom. For meson-meson structure, only the color singlet-singlet $(1\times1)$ is taken into account because of the effect of hidden color channel coupling is considered in QDCSM~\cite{QDCSM1,QDCSM2}.

\begin{table*}[!htb]
\begin{center}
\caption{\label{channels} All possible channels for all quantum numbers}
\begin{tabular}{|ccc|ccc|ccc|cccccc}
\hline\hline
\multicolumn{3}{|c}{$IJ^{P}=00^{+}$} &\multicolumn{3}{|c|}{$IJ^{P}=01^{+}$} &\multicolumn{3}{c|}{$IJ^{P}=02^{+}$} \\
index    &$F^{i}_{I}; S^{j}_{s}; \chi^{c}_{k}$   &channels      & index   &$F^{i}_{I}; S^{j}_{s}; \chi^{c}_{k}$   &channels   & index   &$F^{i}_{I}; S^{j}_{s}; \chi^{c}_{k}$    &channels     \\
\multicolumn{1}{|c}{} &[i;j;k] & \multicolumn{1}{c|}{}  &\multicolumn{1}{c}{} &[i;j;k] & \multicolumn{1}{c|}{} &\multicolumn{1}{c}{} &[i;j;k] & \multicolumn{1}{c|}{} \\ \hline
 1 & [1,1,1] &$\eta_{c}\eta_{s}$     &1 & [1,3,1] &$\eta_{c}\phi$         &1 & [1,6,1] &$J/\psi\phi$\\
 2 & [2,1,1] &$D_{s}\bar{D}_{s}$     &2 & [2,3,1] &$D_{s}{D}_{s}^{*}$     &2 & [2,6,1] &$D_{s}^{*}{D}_{s}^{*}$\\
 3 & [1,2,1] &$J/\psi \phi$          &3 & [1,4,1] &$J/\psi\eta_{s}$       &3 & [3,6,3] &$(cs)(\bar{c}\bar{s})$\\
 4 & [2,2,1] &$D_{s}^{*}\bar{D}_{s}^{*}$ &4 & [2,4,1] &$D_{s}^{*}{D}_{s}$ &4 & [3,6,4] &$(cs)(\bar{c}\bar{s})$\\
 5 & [3,1,3] &$(cs)(\bar{c}\bar{s})$ &5 & [1,5,1] &$J/\psi\phi$           & \multicolumn{3}{c|}{} \\
 6 & [3,1,4] &$(cs)(\bar{c}\bar{s})$ &6 & [2,5,1] &$D_{s}^{*}{D}_{s}^{*}$ & \multicolumn{3}{c|}{}\\
 7 & [3,2,3] &$(cs)(\bar{c}\bar{s})$ &7 & [3,3,3] &$(cs)(\bar{c}\bar{s})$ & \multicolumn{3}{c|}{} \\
 8 & [3,2,4] &$(cs)(\bar{c}\bar{s})$ &8 & [3,3,4] &$(cs)(\bar{c}\bar{s})$ & \multicolumn{3}{c|}{}\\
   \multicolumn{3}{|c|}{}             &9 & [3,4,3] &$(cs)(\bar{c}\bar{s})$ & \multicolumn{3}{c|}{}\\
   \multicolumn{3}{|c|}{}             &10& [3,4,4] &$(cs)(\bar{c}\bar{s})$ & \multicolumn{3}{c|}{} \\
   \multicolumn{3}{|c|}{}             &11& [3,5,3] &$(cs)(\bar{c}\bar{s})$ &  \multicolumn{3}{c|}{}\\
   \multicolumn{3}{|c|}{}             &12& [3,5,4] &$(cs)(\bar{c}\bar{s})$ &  \multicolumn{3}{c|}{}\\
\hline\hline
\end{tabular}
\end{center}
\end{table*}

The energy of $c\bar{c}s\bar{s}$ tetraquark system with $IJ^{P}=00^{+}$, $01^{+}$, and $02^{+}$ for both the meson-meson and diquark-antidiquark structures, as well as the channel coupling of these two structures are listed in Table~\ref{00}, ~\ref{01}, ~\ref{02}, respectively. In those tables, the first column represents the index of every possible channel; the second column lists the corresponding physical channels; the third column indicates the theoretical threshold of every channel; the fourth column ($E_{sc}$) is the energy of every single channel; the fifth column ($E_{cc}$) shows the energy by channel coupling of one certain configuration; the last column ($E_{mix}$) is the lowest energy of the system by coupling all channels of both two configurations.

\begin{table*}[!htb]
\begin{center}
\caption{\label{00} The lowest-lying eigenenergies of $c\bar{c}s\bar{s}$ tetraquarks with $IJ^{P}=00^+$ in the QDCSM. }
\begin{tabular}{ccccccccccccccc}
\hline\hline
~~~~~~Index~~~~~~   ~~~&Channel~~~                           ~~~&Threshold~~~     ~~~&$E_{sc}$~~~  ~~~&$E_{cc}$~~~  ~~~&$E_{mix}$~~~ \\
 ~~~~~~1~~~~~~      ~~~&$\eta_{s\bar{s}}\eta_{c\bar{c}}$~~~  ~~~&3942~~~          ~~~&3944~~~     ~~~&3938~~~       ~~~&3930~~~      \\
 ~~~~~~2~~~~~~      ~~~&$D_{s}\bar{D_{s}}$~~~                ~~~&3936~~~          ~~~&3938~~~     ~~~&~~~                      \\
 ~~~~~~3~~~~~~      ~~~&$J/\psi \phi$~~~                     ~~~&4117~~~          ~~~&4119~~~     ~~~&~~~                     \\
 ~~~~~~4~~~~~~      ~~~&$D_{s}^{*}\bar{D}_{s}^{*}$~~~        ~~~&4224~~~          ~~~&4226~~~     ~~~&~~~                      \\
 ~~~~~~5~~~~~~      ~~~&$(cs)(\bar{c}\bar{s})$~~~            ~~~&~~~              ~~~&4324~~~     ~~~&4219~~~                  \\
 ~~~~~~6~~~~~~      ~~~&$(cs)(\bar{c}\bar{s})$~~~            ~~~&~~~              ~~~&4442~~~     ~~~&~~~                      \\
 ~~~~~~7~~~~~~      ~~~&$(cs)(\bar{c}\bar{s})$~~~            ~~~&~~~              ~~~&4405~~~     ~~~&~~~                      \\
 ~~~~~~8~~~~~~      ~~~&$(cs)(\bar{c}\bar{s})$~~~            ~~~&~~~              ~~~&4305~~~     ~~~&~~~                      \\
\hline\hline
\end{tabular}
\end{center}
\end{table*}

\begin{table*}[!htb]
\begin{center}
\caption{\label{01} The lowest-lying eigenenergies of $c\bar{c}s\bar{s}$ tetraquarks with $IJ^{P}=01^+$ in the QDCSM. }
\begin{tabular}{ccccccccccccccc}
\hline\hline
~~~~~~Index~~~~~~     ~~~&Channel~~~                     ~~~&Threshold~~~     ~~~&$E_{sc}$~~~    ~~~&$E_{cc}$~~~  ~~~&$E_{mix}$~~~ \\
      ~~~~~~1~~~~~~   ~~~&$\eta_{c\bar{c}}\phi$~~~       ~~~&4004~~~           ~~~&4006          ~~~&4006~~~      ~~~&4006~~~      \\
      ~~~~~~2~~~~~~   ~~~&$D_{s}\bar{D_{s}^{*}}$~~~      ~~~&4080~~~           ~~~&4082          ~~~&~~~          ~~~&~~~                \\
      ~~~~~~3~~~~~~   ~~~&$J/\psi \eta_{s\bar{s}}$~~~    ~~~&4055~~~           ~~~&4057          ~~~&~~~          ~~~&~~~           \\
      ~~~~~~4~~~~~~   ~~~&$D_{s}^{*}\bar{D_{s}}$~~~      ~~~&4080~~~           ~~~&4082          ~~~&~~~          ~~~&~~~            \\
      ~~~~~~5~~~~~~   ~~~&$J/\psi \phi$~~~               ~~~&4117~~~           ~~~&4119          ~~~&~~~          ~~~&~~~            \\
      ~~~~~~6~~~~~~   ~~~&$D_{s}^{*}\bar{D}_{s}^{*}$~~~  ~~~&4224~~~           ~~~&4226          ~~~&~~~          ~~~&~~~              \\
      ~~~~~~7~~~~~~   ~~~&$(cs)(\bar{c}\bar{s})$~~~      ~~~&~~~               ~~~&4375          ~~~&4327~~~      ~~~&~~~              \\
      ~~~~~~8~~~~~~   ~~~&$(cs)(\bar{c}\bar{s})$~~~      ~~~&~~~               ~~~&4419          ~~~&~~~          ~~~&~~~              \\
      ~~~~~~9~~~~~~   ~~~&$(cs)(\bar{c}\bar{s})$~~~      ~~~&~~~               ~~~&4375          ~~~&~~~          ~~~&~~~              \\
      ~~~~~~10~~~~~~  ~~~&$(cs)(\bar{c}\bar{s})$~~~      ~~~&~~~               ~~~&4419          ~~~&~~~          ~~~&~~~             \\
      ~~~~~~11~~~~~~  ~~~&$(cs)(\bar{c}\bar{s})$~~~      ~~~&~~~               ~~~&4413          ~~~&~~~          ~~~&~~~              \\
      ~~~~~~12~~~~~~  ~~~&$(cs)(\bar{c}\bar{s})$~~~      ~~~&~~~               ~~~&4352          ~~~&~~~          ~~~&~~~              \\
\hline\hline
\end{tabular}
\end{center}
\end{table*}

\begin{table*}[!htb]
\begin{center}
\caption{\label{02} The lowest-lying eigenenergies of $c\bar{c}s\bar{s}$ tetraquarks with $IJ^{P}=02^+$ in the QDCSM. }
\begin{tabular}{ccccccccccccccc}
\hline\hline
  ~~~~~~Index~~~~~~   ~~~&Channel~~~                      ~~~& Threshold~~~    ~~~& $E_{sc}$~~~  ~~~&$E_{cc}$~~~  ~~~&$E_{mix}$~~~ \\
  ~~~~~~1~~~~~~       ~~~&$J/\psi \phi$~~~                    ~~~&4117~~~       ~~~&4122~~~      ~~~&4121~~~       ~~~&4119~~~       \\
  ~~~~~~2~~~~~~       ~~~&$D_{s}^{*}\bar{D}_{s}^{*}$~~~       ~~~&4224~~~       ~~~&4229~~~      ~~~&~~~           ~~~&~~~             \\
  ~~~~~~3~~~~~~       ~~~&$(cs)(\bar{c}\bar{s})$~~~           ~~~&~~~           ~~~&4429~~~      ~~~&4420~~~       ~~~&~~~               \\
  ~~~~~~4~~~~~~       ~~~&$(cs)(\bar{c}\bar{s})$~~~           ~~~&~~~           ~~~&4437~~~      ~~~&~~~          ~~~&~~~              \\
\hline\hline
\end{tabular}
\end{center}
\end{table*}

The $IJ^P=00^+$ system: Four possible meson-meson channels, $\eta_{s\bar{s}}\eta_{c\bar{c}}, D_{S}\bar{D_{s}}, J/\psi \phi, D_{s}^{*}\bar{D}_{s}^{*}$, and two diquark-antiquark channels, $((cs)(\bar{c}\bar{s}))(3\times\bar{3})$ and $((cs)(\bar{c}\bar{s}))(6\times\bar{6})$, are studied in QDCSM. All results with the $IJ^P=00^+$ are given in Table~\ref{00}. We can see that the energy of every single channel for the meson-meson structure is higher than the corresponding theoretical threshold, which indicates that there is no any bound state. For the diquark-antidiquark configuration, all the masses are higher than the lowest energy of $D_{s}\bar{D_{s}}$ in our model calculation, and the minimum energy is 4305 MeV. Then, we perform the channel-coupling calculation on both the meson-meson and diquark-antidiquark structure, respectively. The energy of the meson-meson structure is 3938 MeV, almost the same as the lowest single channel $D_{s}\bar{D_{s}}$, which indicates that the effect of the channel coupling is quite weak and no bound state is found for the meson-meson structure. For the diquark-antidiquark structure, although the coupling is rather stronger than the meson-meson structure, the energy is still higher than the theoretical threshold of the lowest channel $D_{s}\bar{D_{s}}$.

However, the lowest energy of $3930$ MeV is obtained by coupling all channels of two structures, which is 6 MeV lower than the threshold of the lowest channel $D_{s}\bar{D_{s}}$, which means that there is a bound state for the $IJ^{P}=00^{+}$ $c\bar{c}s\bar{s}$ tetraquark system with mass of $3930$ MeV.
In addition, to explore the structure of this bound state, we calculate the proportion of each channel, and find that the percentage of the $D_{s}\bar{D}_{s}$ state is about $85\%$, while the percentages of the other seven channels are much smaller. This means that the largest contribution to forming this bound state comes from the $D_{s}\bar{D}_{s}$ channel, so this bound state tends to be a molecular state. Moreover, this value is in proximity to the $\chi_{c0}(3930)$ observed by the LHCb collaboration. So we can explain the $\chi_{c0}(3930)$ as a molecular state $D_{s}\bar{D}_{s}$ in present quark model calculation. This result is consistent with the work of Ref.~\cite{lattice_X3930}, in which the lattice QCD calculation with $m_{\pi}\simeq 280$ MeV indicated the existence of the scalar $\bar{D}_{s}D_{s}$ bound state, which might correspond to the $\chi_{c0}(3930)$ observed by the LHCb collaboration~\cite{LHCb_X3930.1,LHCb_X3930.2}. Also, in Ref~\cite{BS_X3930}, two pole positions of $\bar{D}_{s}D_{s}$ system were obtained by solving the Bethe-Salpeter equation, which explained the properties of new exotic resonance $\chi_{c0}(3930)$.

The $IJ^P=01^+$ system: From Table~\ref{channels}, there are six meson-meson channels and six diquark-antidiquark channels. Table~\ref{01} lists the calculated masses of these channels and also their coupling results. The energy range of every single-channel of the meson-meson structure is about $4.0-4.2$ GeV, and the mass of the diquark-antidiquark channel is around 4.4 GeV. All these single channels are unbound.
By coupling the channels with the same configuration, the lowest masses are located at 4006 MeV for the meson-meson structure and 4327 MeV for the diquark-antidiquark structure, both of which are still above the threshold of the lowest channel $\eta_{c\bar{c}}\phi$, indicating that no any bound state exists in the meson-meson structure or the diquark-antidiquark structure. Meanwhile, the lowest energy is still 4006 MeV by the full channel coupling calculation, which means that the effect of all channel coupling is very minor here, and there is no any bound state in the $IJ^P=01^+$ system.

The $IJ^P=02^+$ system: Table~\ref{02} shows that there are two channels ($J/\psi \phi$ and $D_{s}^{*}\bar{D}_{s}^{*}$) of the meson-meson structure and two channels of the diquark-antidiquark configurations for the $IJ^P=02^+$ system. The situation is similar to the $IJ^P=01^+$ case. The energy of each channel is above the threshold of the corresponding channel. Meanwhile, the channel coupling cannot help too much, the lowest energy is still higher than the threshold of the lowest channel $J/\psi \phi$. Therefore, there is no any bound state in the $IJ^P=02^+$ system at present calculation.

\begin{figure*}
\includegraphics[scale=0.35]{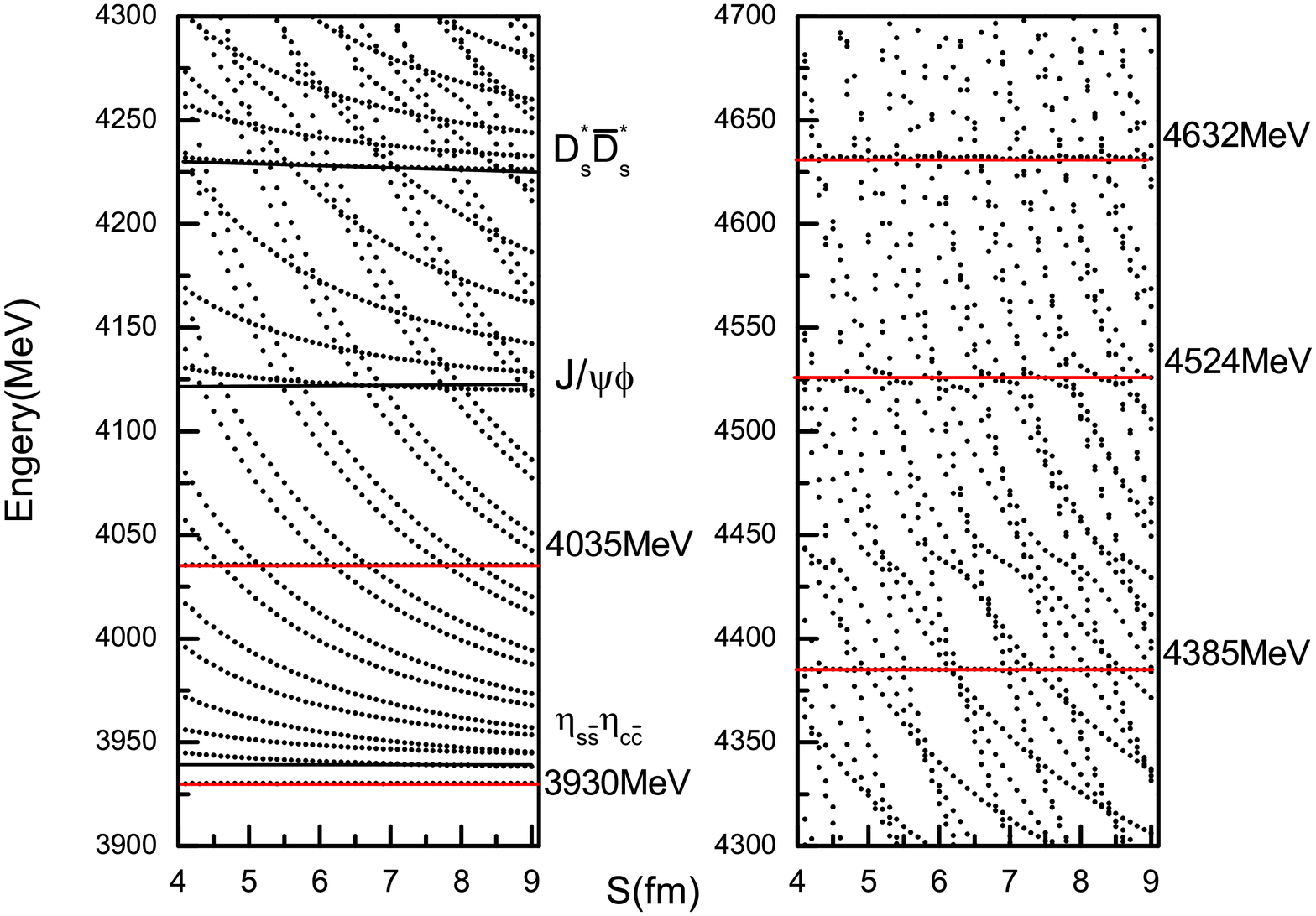}
\vspace{-0.5cm} \caption{The stabilization plots of the energies of the $c\bar{c}s\bar{s}$ with $IJ^{P}=00^+$ in QDCSM.  }
\label{phase00}
\end{figure*}

\begin{figure*}
\includegraphics[scale=0.35]{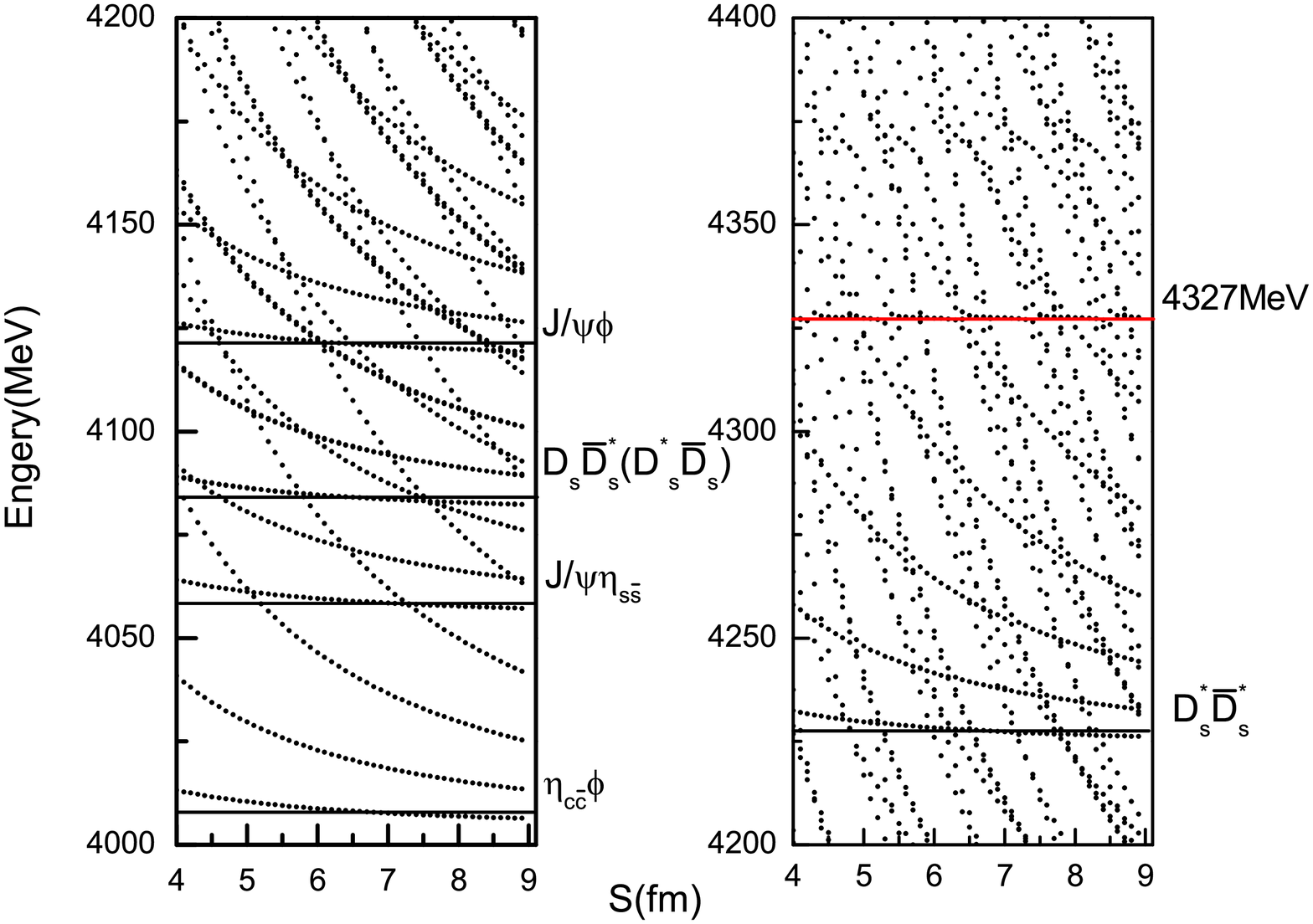}
\vspace{-0.5cm} \caption{The stabilization plots of the energies of the $c\bar{c}s\bar{s}$ with $IJ^{P}=01^+$ in QDCSM. }
\label{phase01}
\end{figure*}

\begin{figure*}
\includegraphics[scale=0.35]{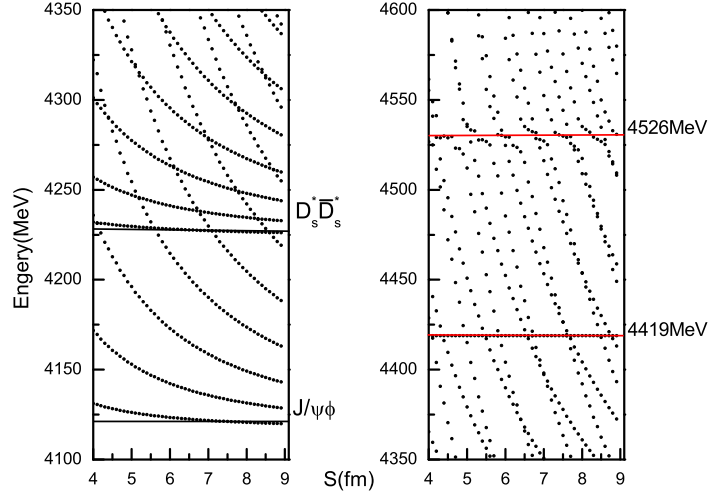}
\vspace{-0.5cm} \caption{The stabilization plots of the energies of the $c\bar{c}s\bar{s}$ with $IJ^{P}=02^+$ in QDCSM. }
\label{phase02}
\end{figure*}

Although there is no any bound state for the $IJ^P=01^+$ and $IJ^P=02^+$ system, some resonance states are still possible in the $c\bar{c}s\bar{s}$ tetraquark system.
The colorful subclusters diquark and antidiquark cannot fall apart directly due to the color confinement, so it is possible for them to be resonance states.
To find out if there is any resonance state, a stabilization method (also named a real scaling method), which has been successfully applied in other multiquark systems~\cite{real_method2, real_method3}, is used in this work. To realize the real scaling method in our calculation, the distance between two clusters is defined as $S$. With the increase of $S$, each state will fall off towards its threshold, except the resonance state, the energy of which will be stable because it will not be affected by the boundary at a large distance. So we calculate the energy eigenvalues of the $c\bar{c}s\bar{s}$ systems by taking the value of $S$ from 4.1 fm to 9.0 fm to see if there is any stable state. The results of the $c\bar{c}s\bar{s}$ tetraquark systems with $IJ^{P}=0{0^+}, 0{1^+}$ and $0{2^+}$ are shown in Fig~\ref{phase00}, Fig~\ref{phase01} and Fig~\ref{phase02}, respectively.

For $IJ^{P}=0{0^+}$ system in Fig~\ref{phase00}, it is obvious that the lowest horizontal line locates at the energy of $3930$ MeV, which represents the bound state of this system. Then, three horizontal lines, which stand for the thresholds of $\eta_{s\bar{s}}\eta_{c\bar{c}}$, $J/\psi \phi$ and $D^{*}_{s}\bar{D}^{*}_{s}$, are marked in Fig~\ref{phase00}.
Besides, another four horizontal lines appear in Fig~\ref{phase00}, corresponding to four resonance states
with the energy around 4035 MeV, 4385 MeV, 4524 MeV, and 4632 MeV, respectively. By comparing with the experimental results, we find that the energy of 4385 MeV is close to the
$X(4350)$, and the quantum number $IJ^{P}=0{0^+}$ is consistent with the reported data by the Belle Collaboration~\cite{Bell-CCSS}. So we explain the $X(4350)$ as a compact tetraquark resonance state with $IJ^{P}=0{0^+}$ in present calculation. Our result is also agrees with the results of the Born-Oppenheimer approach, in which a mass of $4370$ MeV was obtained~\cite{X(4350)CCSS}. Besides, in Ref~\cite{CHQM-CCSS}, $X(4350)$ was also a good candidate of the compact tetraquark state with $IJ^{P}=0{0^+}$ in the chiral quark model.
Similarly, the resonance energy of 4524 MeV is close to the X(4500), and the quantum number $IJ^{P}=0{0^+}$ is also consistent with the reported result of the LHCb Collaboration ~\cite{LHCb-CCSS-2017.2}. So $X(4500)$ is possible to be a compact tetraquark resonance state with $IJ^{P}=0{0^+}$ in present calculation.
In addition to the $X(4350)$ and $X(4500)$, another resonance state with energy around 4632 MeV is obtained. Although the mass is very close to the $X(4630)$, the quantum number $IJ^{P}=0{0^+}$ is different from the reported one $IJ^{P}=0{1^-}$~\cite{LHCb-CCSS-2021}. However, the mass is also close to the state $X(4700)$, and the quantum number is also fit to the experimental data of the LHCb Collaboration~\cite{LHCb-CCSS-2017.2}. So we prefer to assign the resonance state with energy 4632 MeV to be the exotic state $X(4700)$.

For $c\bar{c}s\bar{s}$ system with $IJ^{P}=0{1^+}$ in Fig.~\ref{phase01}, the first six horizontal lines located at the corresponding physical threshold of six channels, which are $\eta_{c\bar{c}}\phi$, $D_{s}\bar{D_{s}^{*}}$, $J/\psi \eta_{s\bar{s}}$, $D_{s}^{*}\bar{D_{s}}$,
$J/\psi \phi$ and $D_{s}^{*}\bar{D}_{s}^{*}$. Obviously, a resonant state is obtained at the energy around 4327 MeV. Although the mass is very close to the $X(4350)$, the quantum number is not quite applicable. However, the LHCb Collaboration claimed the existence of the $X(4274)$ and the measured quantum number was $J^{P}=1^{+}$~\cite{LHCb-CCSS-2017.1}.
Therefore, we tend to use the resonance state around 4327 MeV to explain the $X(4274)$ state in this work. This is similar to the result of Ref~\cite{2011_CCSS}, in which a resonance state with energy near 4.3 GeV is considered as the $X(4274)$.

For the last system $IJ^{P}=0{2^+}$ in Fig.~\ref{phase02}, the first two horizontal lines represent obviously the thresholds of two channels: $J/\psi \phi$ and $D_{s}^{*}\bar{D}_{s}^{*}$. Another two horizontal lines stand for two resonance states, the energy of which is about 4419 MeV and 4526 MeV, respectively.
One may note that the energy of 4526 MeV is also very close to the mass of the $X(4500)$, but the quantum number is not consistent with the experimental data.
So these two resonance state maybe some new exotic states.

\section{Summary}
The $cs\bar{c}\bar{s}$ tetraquark systems with $IJ^{P}=00^{+}$, $01^{+}$ and $02^{+}$ have been systemically investigated by using the RGM in the framework of QDCSM. Our goal is to search for any bound state or resonance state to explain the exotic states, which have been recently observed in the invariant mass distribution of $J/\psi \phi$
and another exotic state $\chi_{c0}(3930)$ observed by the LHCb collaboration.
In this work, two structures: the meson-meson and diquark-antidiquark structures are taken into account. Both single-channel and channel-coupling calculations are performed.
Besides, to search for any resonance state, a stabilization method is applied to the coupling calculation of all channels of both two configurations.

The numerical results show that we obtain a bound molecular state $\bar{D}_{s}D_{s}$ with the quantum number $IJ^{P}=00^{+}$ and the energy 3930 MeV, which can be used to explain the observed $\chi_{c0}(3930)$. Moreover, several resonant states are obtained in this work, which are four $IJ^{P}=00^{+}$ states with the resonance masses around 4035 MeV, 4385 MeV, 4524 MeV, and 4632 MeV, respectively; one $IJ^{P}=01^{+}$ state with the resonance mass around 4327 MeV; and two $IJ^{P}=02^{+}$ states with the resonance masses around 4419 MeV and 4526 MeV, respectively.
All of them are obtained by coupling all channels of both the meson-meson and diquark-antidiquark structures, so they are compact tetraquarks in present quark model calculations.
By comparing with the experimental data, we are inclined to explain the exotic states $X(4350)$, $X(4500)$ and $X(4700)$ as the compact tetraquark state with $IJ^{P}=00^{+}$.
The $X(4274)$ is possible to be a candidate of the compact tetraquark state with $IJ^{P}=01^{+}$.

All these resonance states are worth searching by experiments. We suggest more experimental tests to check the existence of all these possible resonance
states. In addition, to confirm the existence of these $cs\bar{c}\bar{s}$ tetraquark, the study of the scattering
process of the corresponding open channels is needed in future work.

\acknowledgments{This work is supported partly by the National Natural Science Foundation of China under
Contract No. 11675080, No. 11775118, No. 11535005, and No. 11775050.}


\begin{thebibliography}{99}
\bibitem{tr_model1}M. Gell-Mann, Phys. Lett. {\bf8}, 214 (1964).
\bibitem{tr_model2}G. Zweig, in DEVELOPMENTS IN THE QUARK THEORYOF HADRONS. VOL. 1. 1964 - 1978, edited by D. Lichtenberg and S. P. Rosen (1964) pp. 22-101
\bibitem{Bell-X(3872)}S. Choi et al. (Belle), Phys. Rev. Lett.{\bf 91}, 262001 (2003).
\bibitem{Zc(3900)1}M. Ablikim et al. (BESIII), Phys. Rev. Lett. {\bf110}, 252001 (2013).
\bibitem{Zc(3900)2}Z. Liu et al. (Belle), Phys. Rev. Lett. {\bf110}, 252002 (2013).
\bibitem{Zc(3900)3}M. Ablikim et al. (BESIII), Phys. Rev. Lett. {\bf112}, 022001 (2014).
\bibitem{Zc(4020)1}M. Ablikim et al. (BESIII), Phys. Rev. Lett. {\bf112}, 132001 (2014).
\bibitem{Zc(4020)2}M. Ablikim et al. (BESIII), Phys. Rev. Lett. {\bf111}, 242001 (2013).
\bibitem{Zc(3985)}M. Ablikim et al. (BESIII),  Phys. Rev. Lett. {\bf126}, 102001 (2021).
\bibitem{Pc_2019}R. Aaij et al. (LHCb), Phys. Rev. Lett. {\bf122}, 222001 (2019).
\bibitem{LHCb_X3930.1}R. Aaij et al. (LHCb), Phys. Rev. Lett. {\bf125}, 242001 (2020).
\bibitem{LHCb_X3930.2}R. Aaij et al. (LHCb), Phys. Rev. D {\bf102}, 112003 (2020).
\bibitem{CDF-CCSS}T. Aaltonen et al. (CDF Collaboration), Phys. Rev. Lett. {\bf102}, 242002 (2009).
\bibitem{LHCb-CCSS}R. Aaij et al. (LHCb Collaboration), Phys. Rev. D {\bf85}, 091103 (2012).
\bibitem{CMS-CCSS}S. Chatrchyan et al. (CMS Collaboration),  Phys. Lett. B {\bf734}, 261 (2014).
\bibitem{D0-CCSS}V. M. Abazov et al. (D0 Collaboration),  Phys. Rev. D {\bf89}, 012004 (2014).
\bibitem{BABAR-CCSS}J. P. Lees et al. (BABAR Collaboration), Phys. Rev. D {\bf91}, 012003 (2015)
\bibitem{Bell-CCSS}C. P. Shen et al. (Belle Collaboration),  Phys. Rev. Lett. {\bf104}, 112004 (2010).
\bibitem{CDF-CCSS-4274}T. Aaltonen et al. (CDF Collaboration), Mod. Phys. Lett. A {\bf32}, 1750139 (2017).
\bibitem{LHCb-CCSS-2017.1}R. Aaij et al. (LHCb Collaboration), Phys. Rev. Lett. {\bf118}, 022003 (2017).
\bibitem{LHCb-CCSS-2017.2}R. Aaij et al. (LHCb Collaboration), Phys. Rev. D {\bf95}, 012002 (2017).
\bibitem{LHCb-CCSS-2021} R. Aaij et al. (LHCb Collaboration), arXiv:2103.01803v1 (2021).
\bibitem{Canonical-CCSS} P. G. Ortega, J. Segovia, D. R. Entem, and F. Fernandez, Phys. Rev. D {\bf94}, 114018 (2016).
\bibitem{relative-CCSS}Qi-Fang Lv and Yu-Bing Dong, Phys. Rev. D {\bf94}, 074007 (2016).
\bibitem{effect-cups-CCSS}E. S. Swanson, Phys. Rev. D {\bf91}, 034009 (2015).
\bibitem{di-antiquark-CCSS}J. Wu, Y. R. Liu, K. Chen, X. Liu, and S. L. Zhu, Phys. Rev. D {\bf94}, 094031 (2016).
\bibitem{di-antiquark-CCSS1}L. Maiani, A. D. Polosa, and V. Riquer, Phys. Rev. D {\bf94}, 054026 (2016).
\bibitem{Stancu-CCSS}F. Stancu, J. Phys. G {\bf37}, 075017 (2010).
\bibitem{QCD-sum-rule-CCSS}H. X. Chen, E. L. Cui, W. Chen, X. Liu, and S. L. Zhu, Eur. Phys. J. C {\bf77}, 160 (2017).
\bibitem{color-flux-tube-CCSS}C. R. Deng, J. L. Ping, H. X. Huang, and F. Wang, Phys. Rev. D {\bf98}, 014026 (2018).
\bibitem{CHQM-CCSS}Y. F. Yang and J. L. Ping, Phys. Rev. D {\bf99}, 094032 (2018).
\bibitem{wang-diquark-CCSS1}Z. G. Wang, Eur. Phys. J. C  {\bf77}, 78 (2017).
\bibitem{wang-diquark-CCSS2}Z. G. Wang, Eur. Phys. J. C  {\bf76}, 657 (2016).
\bibitem{rescattering-CCSS} X. H. Liu,  Phys. Lett. B {\bf766} 117 (2017).
\bibitem{Ebert} D. Ebert, R. N. Faustov, V.O. Galkin, Eur. Phys. J. C {\bf58}, 399 (2008).
\bibitem{QDCSM_explain1}F. Wang, G.H. Wu, L.J. Teng, J.T. Goldman, Phys. Rev. Lett. {\bf69}, 2901 (1992).
\bibitem{QDCSM_explain2}L.Z. Chen, H.R. Pang, H.X. Huang, J.L. Ping, F. Wang, Phys. Rev. C {\bf76}, 014001 (2007).
\bibitem{QDCSM_explain3}M. Chen, H.X. Huang, J.L. Ping, F.Wang, Phys. Rev. C {\bf83}, 015202 (2011).
\bibitem{native}A. De Rujula, H. Georgi, and S. L. Glashow, Phys. Rev. D 12, 147 (1975); N.Isgur andG.Karl, ibid. {\bf18}, 4187 (1978); {\bf19}, 2653 (1979); {\bf20}, 1191 (1979).
\bibitem{QDCSM1}G. H. Wu, L. J. Teng, J. L. Ping, F. Wang, and T. Goldman, Phys. Rev. C {\bf53}, 1161 (1996).
\bibitem{QDCSM2} H. X. Huang, P. Xu, J. L. Ping, and F. Wang, Phys. Rev. C {\bf84}, 064001 (2011).
\bibitem{Pc_huang1}H. X. Huang, Ch. R. Deng, J. L. Ping, F. Wang, Eur. Phys. J. C {\bf76}, 624(2016).
\bibitem{PDG}M. Tanabashi et al. (Particle Data Group), Phys. Rev. D {\bf 98}, 030001 (2018).
\bibitem{si_1}G. H. Wu, J. L. Ping, L. J. Teng, F. Wang, and T. Goldman, Nucl. Phys. A {\bf673}, 279 (2000).
\bibitem{si_2}J. L. Ping, F. Wang, and T. Goldman, Nucl. Phys. A {\bf657}, 95 (1999).
\bibitem{si_3}H. R. Pang, J. L. Ping, F. Wang, and T. Goldman, Phys. Rev. C {\bf65}, 014003 (2001).
\bibitem{si_4}M. M. Xu, M.Yu, and L.S.Liu, Phys. Rev. Lett. {\bf100}, 092301 (2008).
\bibitem{RGM}M. Kamimura, Suppl. Prog. Theor. Phys. {\bf62}, 236 (1977).
\bibitem{lattice_X3930}S. Prelovsek, S. Collins, D. Mohler, M. Padmanath, and S. Piemonte, (2020), arXiv:2011.02542.
\bibitem{BS_X3930}Xiang-Kun Dong, Feng-Kun Guo, and Bing-Song Zou, Progr. Phys {\bf41}, 65-93 (2021).
\bibitem{real_method1}J.Simon, J. Chen, Phys. {\bf75}, 2465 (1981).
\bibitem{real_method2}E. Hiyama, M. Kamimura, A. Hosaka, H. Toki, and M. Yahiro, Phys. Lett. B {\bf633}, 237-244 (2006).
\bibitem{real_method3}E. Hiyama, A. Hosaka, M. Oka, and J. M. Richard, Phys. Rev. C {\bf98}, no.4, 045208 (2018).
\bibitem{X(4350)CCSS}E. Braaten, C. Langmack, and D. H. Smith, Phys. Rev. D {\bf90}, 014044 (2014).
\bibitem{2011_CCSS}W. Chen, S.L. Zhu, Phys. Rev. D {\bf83}, 034010 (2011).
\end{thebibliography}
\end{document}